\documentstyle[12pt,psfig]{article}
\begin{document}
\baselineskip 6mm
{\noindent  \large \bf 
MOMENTUM DENSITY OF HIGH T$_c$ COPPER OXIDES
\vspace{6mm}}

\noindent{\large
B. Barbiellini and P.M. Platzman }
\vspace{2mm}

{\small
\noindent
Bell Laboratories, Lucent Technologies, 700 Mountain Ave,
Murray Hill, NJ 07974, USA
\vspace{2mm}
\vspace{6mm}

\noindent
{\bf \underline{Keywords:}} 
high $T_c$ superconductors,
Compton profile,
positron annihilation spectroscopy.
\vspace{3cm}
  }                 %small

\noindent{\bf Abstract:}
{\small We discuss the work aimed at studying the
momentum density of high $T_c$ superconducting oxides
using Compton profile and positron annihilation
spectroscopies. 
         }  
\vspace{4mm}

Several experiments and theoretical models suggest that strong electron
correlation effects are involved in the description 
of copper oxides \cite{dagotto}.
While searching for the Fermi surface (FS), 
a more general question to ask is: What 
is the influence of strong correlations
on the momentum density ?
Unfortunately, the overall momentum density in
such materials is described quite well by a local
tight-binding description of the transition
metal surrounded by, for example, an octahedron of 
oxygen atoms \cite{chiba}. Thus the dominant
features are due to wavefunctions effects,
and the FS plays a minor role.
Instructive examples of the importance of orbital contributions in 
momentum density are found from the 2D-ACAR 
(2-Dimensional Angular Correlation of the Annihilation Radiation) 
technique applied to copper oxides \cite{review}.
The momentum density anisotropies of $\rm La_{2-x}Sr_xCuO_4$ 
\cite{turchi,sterne}
and  $\rm Tl_{2-x}Ba_xCuO_6$  \cite{tbco}
can be described reasonably well by a 
simple LCAO-MO method \cite{chiba}, which neglects the FS. 
Therefore, in these cases the chemical bonding
overshadows the smaller FS effects and it is very difficult
to extract the FS signal. 
However, a more favorable case is provided 
by the $\rm YBa_2Cu_3O_{7-\delta}$,
where the 1-dimensional ridge FS has a two-fold symmetry 
which distinguishes it from important four-fold symmetry wave 
function effects \cite{haghighi}.
Another important result obtained by the positron annihilation
spectroscopy \cite{pbco}
is the observation of a ridge FS signal
in the insulating $\rm PrBa_{2}Cu_3O_{7-\delta}$.
Therefore the insulating character of
this compound can be explained by defects in 
the Cu-O chains, that prevent it from conducting,
and by the existence a mechanism which binds
the doped holes to the Pr sites, making the Cu-O
planes insulating \cite{fehrenbacher}.

Several recent publications have suggested that
the unusual features of the $c$-axis resistivities
observed in the cuprates could be an indication
of the non-Fermi-liquid nature 
of these materials \cite{clarke}.
For instance, the $c$-axis resistivities estimated
from band theory are at least
a factor of ten smaller than the experimental values.
Moreover the standard band theory 
for 2-layer cuprates compounds \cite{massidda,pickett}
predicts due to bonding antibonding states 
a two piece FS, while photoemission results in 
$\rm Bi_2Sr_2CaCu_2O_8$
do not show split Fermi surfaces \cite{photo}.
However, the photoemission results do not yet rule
out the Fermi-liquid theory, since they 
might also be explained by strong polarization selections rules. 
 
Anderson, Clarke and coworkers \cite{clarke,anderson,chakravarty}
have suggested that coherent tunneling between 
the planes is forbidden
because of an in-plane Fermi-liquid breakdown 
and that coherent pair tunneling becomes allowed when 
we enter the superconducting state.
This argument remains controversial.
We propose to look at the consequences of this
assumption on the momentum density.

While positron annihilation is a possible technique,
it presents some difficulties.
In copper oxides the spatial distribution of the positron
wave function has a large contribution on the 2D-ACAR spectra.
For instance, in $\rm YBa_2Cu_3O_{7-\delta}$ the positrons 
are mainly located along the Cu-O chains, while in
$\rm Nd_{2-x}Ce_xCuO_4$ they have more overlap with
the Cu-O planes. Therefore, $\rm Nd_{2-x}Ce_xCuO_4$ 
is a more suitable compound for probing the Cu-O planes with
positrons \cite{stanford}. 
Unfortunately the defects contained in the real 
samples make this investigation difficult, since they
trap the positrons.
Nevertheless a recent 2D-ACAR experiment \cite{ndco} indicates
evidence of a FS signal from the Cu-O layers.

Compared with positron annihilation, Compton scattering 
\cite{compton} offers some advantages. 
The results are not sensitive to sample
purity or to lattice defects. Moreover the interpretation of the 
data is not complicated by the positron wave function
and correlation effects. Nevertheless, 
the 2D-ACAR spectrum gives a projection of the momentum density 
along the direction perpendicular to the 2-dimensional detectors,
whereas the Compton profile allows one only to identify structures in 
the momentum density averaged over the planes perpendicular to the 
scattering vector. Therefore, we can view these two spectroscopies
as being complementary tools \cite{bansil} to study the different 
pieces of the FS of the high $T_c$ superconducting oxides.
In the case of $\rm YBa_{2}Cu_3O_{7-\delta}$, 
the 2D-ACAR has identified
the ridge FS associated with the Cu-O chains, while  
the Compton scattering could be used for the study
of the barrel portions of the FS, which are
connected with the interlayer coupling.

The standard local density 
approximation (LDA) \cite{massidda} predicts an energy splitting 
for 2-layer high $T_c$ compounds, due to electron coherent motion 
between the layers.
In $\rm YBa_2Cu_3O_7 $,
where two neighboring Cu-O planes are separated by
$d=6.43$ (a.u.), the coherent motion between Cu-O planes would give 
a feature at $p=2\pi/d$ in the Compton profile along the $c$ axis.  
To detect this feature, one can consider differences
of the directional Compton profiles along two directions:
in plane and along the $c$-axis \cite{grenoble}.
The LDA calculations give a structure
in the differences of the directional Compton profiles
[001]-[100] and [001]-[010] located at $p=2\pi/d \approx  1$ (a.u.)
(see Fig.1).
We have simulated the effect of suppressed tunneling
by shifting the Fermi energy an amount comparable to
the energy separation of bonding and antibonding states.
%*******************************************************************
\begin{figure}
%\vspace{70mm} 
% -------------------------figure en postscript-------------------------------
%\special{psfile=fig1.ps}  
\centerline{\psfig{file=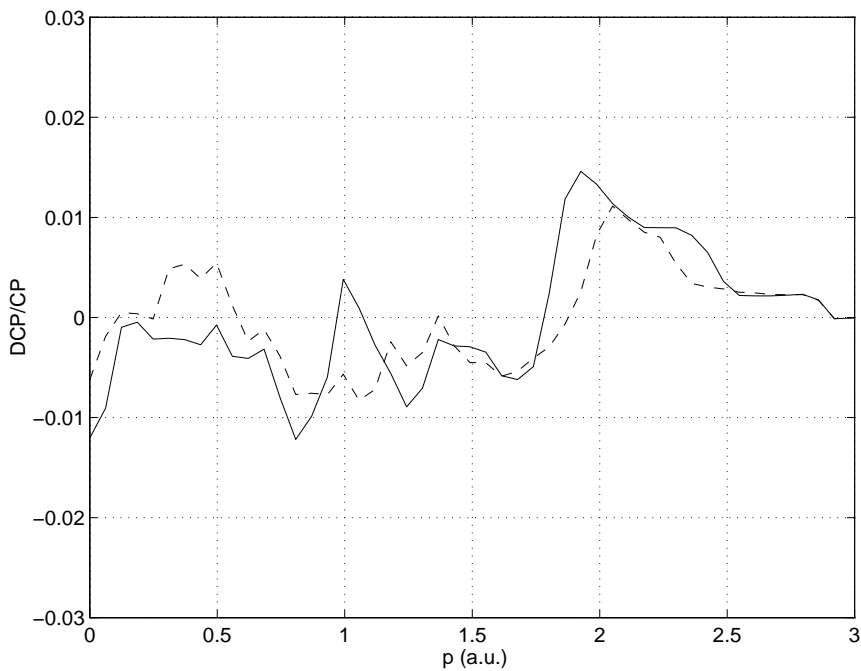,width=4.5in}}
% -------------------------figure en postscript-------------------------------
\caption{Directional differences of Compton profiles [001]-[100]
for twinned  $\rm YBa_2Cu_3O_7$
as given by a band structure calculation. Full line: the peak
near 1 a.u. is the structure mentioned in the text. Dashed line:
the peak disappears by shifting the Fermi Energy by + 1.3 eV. }
\label{fig1}
\end{figure}  
%*******************************************************************
%
The amplitude of this modulation depends on the position of
the Fermi level (or on the difference of electrons 
contained in the two Fermi surfaces of the Cu-O planes) 
and its predicted value is of the order of $1 \% $.
Therefore it should be detectable with more than $10^4$
counts in a resolution bin of $0.15$ (a.u.). 
On the other hand, the absence of this signal, 
which implies a degenerate FS of the Cu-O planes, 
would be consistent with Anderson's suggestion.
  
A very important question regarding the nature
of the electronic ground state of high $T_C$
materials is the character of the extra holes
introduced when we dope away from half filling.
Experiments seem to show that
the holes introduced by doping 
reside primarily on oxygen sites.
The suggestion of Zhang and Rice \cite{zhang}
is that they reside in a molecular orbital state
$P_{MO}=P_{1x}-P_{2y}-P_{3x}+P_{4y}$.
The orbital $P_{MO}$ couples 
with the Cu $3d_{x^2-y^2}$ state and forms a singlet 
named after Zhang and Rice \cite{zhang}.

The difference of the directional 
Compton profiles [100]-[110] should reflect anisotropy
of the Cu ($3d_{x^2-y^2}$) holes (see Fig. 2A).
$P_{MO}$ gives an additional anisotropy.
Its contribution (see Fig. 2B) is expected 
to increase with the doping.
For an optimally doped sample, we expect there to
be a 20-10 \% increase of the anisotropy 
and an additional periodic structure which can
be directly related to the phases of the 
wave function associated with the four oxygens 
surrounding a given Cu atom.
%
%*******************************************************************
\begin{figure}
%\vspace{70mm} 
% -------------------------figure en postscript-------------------------------
%\special{psfile=fig2.ps}
\centerline{\psfig{file=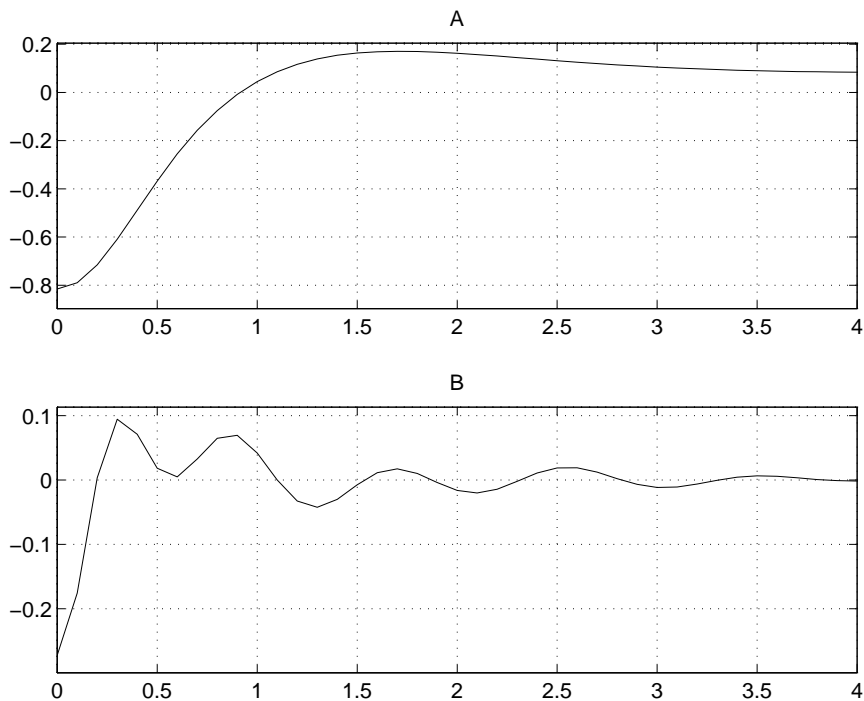,width=4.5in}} 
% -------------------------figure en postscript-------------------------------
\caption{Directional differences of Compton profiles [100]-[001].
A: contribution of the Cu ($3d_{x^2-y^2}$) hole (arbitrary units).
B: contribution of the $P_{MO}$ hole for optimally doped sample.}
\label{fig2}
\end{figure}  
%*******************************************************************
%
However, positron annihilation experiments in
$\rm La_{2-x}Sr_xCuO_4$ \cite{howell} have not
observed significant change in the shape of the anisotropies
as a function of $x$.
One possible explanation is that the Zhang-Rice singlet 
is a positive object which repels the positron. 
If so the Zhang-Rice singlets would not be seen by positron
but may be observed in Compton scattering experiments.

In conclusion 2D-ACAR and Compton scattering high resolution
momentum experiments are useful tools to probe 
the electronic structure of the high $T_c$ superconducting oxides. 
By using these two spectroscopies we can gain insight into the 
question of the electron-electron correlations and of the FS.
%The present experimental resolution is in principle
%sufficient to provide direct measurements of the
%superconducting gap \cite{review,compton}.

{\noindent {\bf Acknowledgements:}}
We wish to thank A.A. Manuel, A. Shukla, T. Jarlborg, 
F.C. Zhang and E. Isaacs for discussions.
B.B. was supported by the Swiss National Science
Foundation Grant No. 8220-037167.

\end{document}